\documentstyle[12pt]{article}
\def\bc{\begin{center}}           \def\ec{\end{center}}
\def\beq{\begin{equation}}         \def\eeq{\end{equation}}
\def\bear{\begin{eqnarray}}       \def\eear{\end{eqnarray}}
\def\bt{\begin{tabular}}          \def\et{\end{tabular}}
\def\la{\langle}                  \def\ra{\rangle}
\def\lg{\large}                   
\def\lt{\left}                    \def\rt{\right}
\def\dg{\dagger}                  \def\ci{\cite}
\def\lb{\label}                   \def\ld{\ldots}
\def\pr{\prime}                   \def\sm{\small}
              \def\rar{\rightarrow}
        
\def\td{\tilde}                   \def\pr{\prime}
\def\pd{\partial}                 
\def\ns{\normalsize}              \def\noi{\noindent}

\def\alf{\alpha}         \def\gam{\gamma}
     
\def\Dlt{\Delta}     \def\dlt{\delta}    
\def\lam{\lambda}    \def\Lam{\Lambda}    \def\sig{\sigma}

\def\matb[#1#2#3#4]{\lt(\bt{ll}$#1$&$#2$\\$#3$&$#4$\et\rt)}
\def\mata[#1#2]{\lt(\bt{l}$#1$\\$#2$\et\rt)}
\def\c#1{{\cal #1}}
\begin{document}

\title{\vspace{-3cm}
{\flushleft{\ns quant-ph/0012044 \\
{\sm To appear in the Proceedings of the \\ 
IInd Int. Conference on Geometry, Integrability\\[-4mm]
and Quantization, Varna, June 2000}}\\[7mm]}
{\bf Diagonalization of Hamiltonians, uncertainty matrices and Robertson
    inequality}}
\author{D.A.  Trifonov\\
        Institute for Nuclear Research and Nuclear Energetics\\
        72 Tzarigradsko chauss\'ee, Sofia, Bulgaria }
\maketitle

\begin{abstract}
The problem of diagonalization of Hamiltonians of $N$-dimensional boson
systems by means of time-dependent canonical transformations (CT) is
considered, the case of quadratic Hamiltonians being treated in greater
detail.  The unitary generator of time-dependent CT which can transform
any Hamiltonian to that of a system of uncoupled stationary oscillators is
constructed.  The close relationship between methods of canonical
transformations, time-dependent integrals of motion and dynamical symmetry
is noted.

The diagonalization and symplectic properties of the uncertainty matrix
for $2N$ canonical observables are studied. It is shown that the normalized
uncertainty matrix is symplectic for the squeezed multimode Glauber
coherent states and for the squeezed Fock states with equal photon numbers
in each mode. The Robertson uncertainty relation for the dispersion matrix
of canonical observables is shown to be minimized in squeezed coherent
states only.
\end{abstract}

\section{Introduction}

The method of  canonical  transformations  (CT)  proved  to  be  a
fruitful approach in treating quantum systems. It is  most  efficient for
systems  that  are  described  by   Hamiltonians, that are quadratic
in coordinates  and  moments,  or  equivalently  in  boson creation  and
annihilation operators (quadratic Hamiltonians).  The  main advantage of the
method of CT consists in reducing the Hamiltonian $H$  of the treated
system $\c S$ to a Hamiltonian $H^\pr $ of some simple system $\c S^\pr$
with known solutions. The well known  example (and probably the first one)
of such an application is the diagonalization of the modeled quadratic
Hamiltonians in superfluidity and superconductivity theory by means of
linear time-independent transformations of boson or fermion operators (the
Bogolyubov transforms) \ci{NNB}. In \ci{MMT1} time-dependent CT for
quadratic systems were used (probably for the first time) in construction
of integrals of motion that are linear in coordinates and moments.

Quadratic Hamiltonians model many quantum (and classical) systems: from
free particle and free electromagnetic  field  to  the waves in nonlinear
media,  molecular  dynamics  and  gravitational waveguide
\ci{MM,SS,Littl,DMM}.  A considerable attention to quadratic
classical and/or quantum systems is paid in the literature for a long
period  of time (see, for example, \ci{MMT2,MM,DMM,Selez,T99} and
references therein).

Diagonalization problem of quadratic Hamiltonians is considered in a
number of papers \ci{Wolf,TG,Tiko,Bogdan,Colpa,Selez}. In general,
quadratic Hamiltonians can not be diagonalized by means of
time-independent CT, even in the one-dimensional case \ci{TG,Bogdan}. In
the one-dimensional case the term proportional to the product of
coordinate and moment can be eliminated by time-dependent CT only. For
this purpose a time-dependent point transformation (i.e. scale or squeeze
transformation) is sufficient \ci{TG}. Time-dependent CT are very
powerful. Seleznyova \ci{Selez} has shown that the Hamiltonian of a
nonstationary quantum oscillator can always be brought to the diagonal
form of that of the stationary harmonic oscillator by means of linear
time-dependent  CT.

The aim  of the  present  paper  is  to  establish the canonical
equivalence  of $N$--dimensional  quantum  systems and to perform it
explicitly in the  case  of systems with quadratic Hamiltonians. Two
systems are called canonically equivalent if their Hamiltonians can be
related by means of a CT. Due to the known fon Neumann theorem CT in
quantum mechanics are generated by unitary operators.  Therefore canonical
equivalence is in fact unitary one.  A second aim of the present paper is
to consider the symplectic properties of the uncertainty matrix for
canonical observables and its diagonalization using linear CT
\ci{SCB,T95}.

The paper is organized as follows.  In  section  II we show   that  any
two $N$--dimensional quantum Hamiltonians (time-dependent, in general)
$H(t)$ and $H^\pr(t)$ can be canonically  related  via  time-dependent
unitary operator $U(t)$. The group of CT which leave $H$ invariant (i.e.
$H = H^\pr$) is shown to coincide with the dynamical symmetry group of the
system. In the case  of two quadratic Hamiltonians the operator $U(t)$ is
an exponent of a quadratic form of coordinates and moments (that is, an
element the methaplectic  group $Mp(N,R)$).  In particular, such operators
can diagonalize any quadratic Hamiltonian. We note that there are two
types of diagonalizations depending of the type of the canonical variables
in which the target Hamiltonian is diagonal.

In section  III we perform  the  diagonalization  of $N$--dimensional
quadratic Hamiltonian, expressing the parameters of the corresponding
linear CT in terms of solutions of  linear  first  order differential
equations.  For $N=1$ these equations are reduced to the equation
$\ddot{z} + \Omega^2(t)z=0$ of classical oscillator with varying
frequency. The relation of CT to the linear integrals of motion is briefly
discussed.

In section IV the main properties of the uncertainty matrix {\lg$\sig$}
for $N$ observables are considered. It is shown that for canonical
observables the uncertainty matrix is positive definite and thus (due to
the known theorem by Williamson
\ci{Will,Tiko}) can be diagonalized by means of linear CT.  For squeezed
canonical coherent states (CS) \ci{SS} and for squeezed Fock states with
equal boson/photon numbers in every mode the  matrix {\lg$\sig$} (when
normalized to unity) is found to be symplectic itself.  The symplectic
character of the normalized uncertainty matrix in squeezed CS can also be
inferred from the results of paper \ci{Simon}.

\section{Unitary equivalence of quantum systems}

The main aim in the method of CT is to reduce the Hamiltonian $H$ of the
treated system $\c S$ to a Hamiltonian $H^\pr $  of some simple system $\c
S^\pr$ with known solutions. CT in quantum theory are generated by unitary
operators $U$, which is called the generator of CT.  If CT is {\it
time-independent} then $H$ and $H^\pr $ are unitary equivalent and their
spectrums are the same. However not any pair $H$ and $H^\pr$ can be
related by means of time-independent CT. In particular, not any quadratic
Hamiltonian can be reduced to that of a harmonic oscillator by means of
time-independent CT \ci{TG,Tiko,Bogdan,Colpa}, even in the one-dimensional
case \ci{TG}. The {\it time-dependent} CT are much more powerful as we
shall see below.

Let $|\Psi (t)\ra$ be a solution of the Schr\"odinger equation $[i\hbar
\pd/\pd t - H]|\Psi (t)\ra = 0.$ Then for any unitary operator $U(t)$ the
transformed state $|\Psi ^\pr (t)\ra$,
$|\Psi ^\pr (t)\ra = U(t)|\Psi (t)\ra$,
obeys the equation $[i\hbar \pd/\pd t - H^\pr]
|\Psi ^\pr (t)\ra = 0$ with the  new  Hamiltonian $H^\pr $,

\beq\lb{H'}    
H^\pr  = U(t)HU^{\dg}(t) - i\hbar U(t)\pd U^{\dg}(t)/{\pd t}.
\eeq
 Conversely, if two Hamiltonians $H$ and $H^\pr $ are related
by  means  of  an (unitary) operator $U(t)$  in  accordance  with
eq. (\ref{H'})  then any solution $|\Psi (t)\ra$ of the system $\c S$ 
is mapped
into a solution $|\Psi ^\pr (t)\ra$  of the system $\c S^\pr $.
However, not any two given solutions $|\Psi (t)\ra$ and $|\Psi
^\pr (t)\ra$ of the two systems could be mapped into each other by  means
of $U(t)$ since $U(t)$ in general cannot act transitively in the Hilbert
space. A more compact form of relation (\ref{H'}) is $D^\pr(t) =
U(t)D(t)U^\dg(t)$, where $D(t) = i\hbar\pd/\pd t - H(t)$. $U(t)$ is
intertwining operator for $D(t)$ and $D^\pr(t)$. When $D(t)$ and
$D^\pr(t)$ act in the same Hilbert space one says that $D(t)$ and
$D^\pr(t)$ are unitary equivalent. $D(t)$ is often called Schr\"odinger
operator.

 From the requirement for the mean values of the  "old"  operator
$A$ and the "new" one $A^\pr $,
$$
\la \Psi (t)\lt|\matrix{A}\rt|\Psi (t)\ra = \la \Psi ^\pr
(t)\lt|\matrix{A^\pr }\rt|\Psi ^\pr (t)\ra,
$$
it follows that the operators A and $A^\pr $ are related as
$A^\pr  = U(t)AU^{\dg}(t)$.
Therefore the new canonical operators of the
coordinates  and moments $q_k^\pr $ and $p_k^\pr , k =1,\ld ,N$
are related to the old ones as
\beq\lb{2b}
q_k^\pr  = U(t)q_{k}U^{\dg}(t),\quad p_k^\pr  = U(t)p_{k}U^{\dg}(t).
\eeq

Two quantum systems should  be  called  {\it canonically} or {\it unitary
equivalent}  if their Schr\"odinger operators are unitary equivalent.
The corresponding Hamiltonian
operators $H$ and $H^\pr $, related in  accordance  with  eq. (\ref{H'}),
should be called {\it canonically equivalent} with respect to
$U(t)$.
Let us note the main three advantages  of establishing unitary equivalence
of two systems (see also \ci{Selez}, where in fact canonical equivalence
of one dimensional oscillators with constant and time-dependent
frequencies was considered):

(a) If we know solutions $|\Psi\ra$ for one of the two  canonically  related
systems we can obtain solutions for the other one as $U(t)|\Psi\ra$.

(b) If a time-dependent state $|\Psi ^\pr (t)\ra$ of the  system $\c S^\pr
$  is  an eigenstate  of  an  operator $A^\pr $  then  its
$U(t)$--partner $|\Psi (t)\ra = U^{\dg}(t)|\Psi ^\pr (t)\ra$ in the system
$\c S$ is an eigenstate of the operator $A = U^{\dg}(t)A^\pr U(t)$.

(c) If the operator $A^\pr $ is an integral of motion  for $\c S^\pr $,
i.e. if $A^\pr$ commutes with the Schr\"odinger operator, $\pd A^\pr /\pd t
- (i/\hbar )[A^\pr ,H^\pr ] = 0$, then the operator
$A = U^{\dg}(t)A^\pr U(t)$
is an integral of motion for the old system $\c S$,

\beq\lb{3a}
\pd A/\pd t - (i/\hbar)[A,H] = 0.
\eeq
This property is very important  since if we know one solution for a given
system $\c S$  we  can  construct new solutions acting by the invariant
operators on the known solution.\\

{\sm\bf Proposition 1}. {\it Any two $N$--dimensional quantum Hamiltonians
$H$  and $H^\pr $ are canonically equivalent. The unitary operator $U(t)$,
that relates $H$ and $H^\pr $ takes the form
\beq\lb{o4}
U(t) =\hbox{ T\,exp}\lt[-\frac{i}{\hbar}\int_{t_0}^{t}
H^\pr(t)dt\rt]\,U_{0}\tilde{\rm T}\,\exp\lt[{i\over
\hbar }\int_{t_0}^t H(t)dt\rt] \equiv
S^\pr(t)U_{0}S^\dg(t),
\eeq
where $U_{0}$  is  constant  unitary  operator  and $T$  and
$\tilde{T}$  stand  for chronological and  antichronological  product.
The  solution  (\ref{o4})  is unique for any initial condition
$U(0) = U_{0}$}.\\[-3mm]

{\it Proof}. Let us perform two successive time-dependent CT by means of
$U_1=U_{0}S^\dg(t)$ and $U_2=S^\pr(t)$,
\beq\lb{o4a}
S^\dg(t) =\td{\rm T}\exp\lt[{i\over \hbar}\int^{t}H(t)dt\rt].
\eeq
 Then from eq. (\ref{H'}) (taking into account $\pd U_1^\dg(t)/\pd t =
(-i/\hbar)HU_1^\dg$) we easily  get $H_{1} = 0$  for  any $U_{0}$.  The
second transformation by means of $U_2=S^\pr(t)$,

\beq\lb{04b}
S^\pr(t) = {\rm T}\exp\lt[-{i\over \hbar}\int^t H^\pr(t)dt\rt],
\eeq
then yields the required result $\lt(\pd U^\dg_2/\pd t = \pd S^{\pr\dg}/\pd
t=(i/\hbar) S^\pr H^\pr\rt)$:

\beq\lb{o4c}
H_{2} = U_2H_{1}U^{\dg}_2 - i\hbar U_{2}\pd U^{\dg}_{2}/\pd t =
-i\hbar S^\pr\pd S^{\pr\,\dg}/\pd t = H^\pr .
\eeq
Now we see that the direct CT: $H \rar  H^\pr $ is performed by the
unitary operator (\ref{o4}).

For a given $H$ and $H^\pr$ the intertwining operator $U(t)$ is not unique.
However the time-dependence of $U(t)$ is uniquely determined
by any initial condition $U(0) = U_{0}$.  Indeed, suppose there is another
unitary operator $\tilde{U}(t)$, which also relates $H$ and $H^\pr $
canonically and $\tilde{U}(0) = U_{0}$.  Now we  note  that  (it is easily
derived from (\ref{H'})) if $\tilde{U}$ transforms $H$ into $H^\pr $ then
$\tilde{U}^{\dg}$ transforms $H^\pr $ back into $H$ and therefor the
product $V \equiv \tilde{U}^{\dg}U$ keeps $H$ invariant:
$$
H = VHV^{\dg} + {i\over \hbar}[\pd V/\pd t] V^{\dg}\quad  {\rm
and}\quad V(0) = 1.
$$
On the other hand, by using eq. (\ref{H'}) for $U$ and $\td{U}$, one
obtains the equality $\pd V/\pd t - (i/\hbar)[V,H] = 0$,
which means that $V$ is an integral of motion for the system
$\c S$. Any invariant operator for $H$ has the  form (note that $S(t)$ is
the evolution operator for $\c S$) $V(t) = S(t)V(0)S^\dg(t)$,
and since $V(0) = 1$ we have $V(t) = \tilde{U}^{\dg}U  = S(t)S^\dg(t) =
1$. In a similar way one can get $U(t)\tilde{U}^{\dg}(t) = 1$. And if
$\tilde{U}^{\dg}U = 1 = U\tilde{U}^{\dg}$, then $U = \tilde{U}$ (because
of the uniqueness of the inverse $U^{-1}$).  End of the proof.

Let us note that not necessarily $H(0) = H^\pr (0)$: we  have

$$ H^\pr (0) = U_{0}HU^{\dg}_{0} -
i\hbar U(0)[\pd U^{\dg}/\pd t]_{|t=0}.$$

Suppose now that $H(t)$ and $H^\pr(t)$ are elements of a Lie algebra $\c
L$.  Then $S \in G \ni S^\pr$, where $G$ is the Lie group generated by
$\c L$.
Thus, the CT generator $U(t)\in G$ (for $U_0 = 1$ and for $U_0\in G$ as
well) and one can use the known properties of $G$ to represent $U(t)$ in
other factorized forms.

The operator (\ref{o4}) converts canonically any $N$--dimensional $H$ into
any desired $N$--dimensional $H^\pr $. In particular $H$ can be converted
into $H^\pr$ for a system of $N$ free particles or for a  system  of
uncoupled  harmonic oscillators ($N$--mode free boson field). In the
latter case  if $H$  is  a quadratic form in terms  of $N$  canonical
operators $q_{k}$  and $p_{j}$  the operator  (\ref{o4})  solves  the
{\it diagonalization problem} for quadratic Hamiltonians.

A CT will be called diagonalizing if the new Hamiltonian $H^\pr$ in terms
of the coordinates and moments is diagonal quadratic form with {\it
constant} coefficients, i.e. $H^\pr $ is a  Hamiltonian  for  a system  of
uncoupled harmonic oscillators $H_{\rm ho}$. One has to distinguish
between two different kinds of diagonalization of $H$:\\[-3mm]

{\it First kind diagonalization}:\,\, $H^\pr $ is diagonal in terms of the
new variables $q_j^\pr , p_k^\pr $,

{\it Second kind diagonalization}:\,\,
$H^\pr $ is diagonal  in  terms of the old variables
$q_{j}, p_{k}$.\\[-3mm]

\noi In the  first  case  the two  systems $\c S$  and $\c S^\pr $
are treated in two different ($q$-  and $q^\pr$- ) coordinate
representations (wave functions $\Psi (q,t) = \la q|\Psi(t) \ra$ and
$\Psi^\pr(q^\pr,t) = \la q^\pr |\Psi^\pr(t)\ra)$,  whereas  in  the second
case one can work in the same $q$-representation  (wave  functions $\Psi
(q,t) = \la q|\Psi (t)\ra$ and $\Psi ^\pr (q,t) = \la q|\Psi ^\pr
(t)\ra$).

The second kind diagonalization is achieved by means of operator $U(t)$,
eq. (\ref{o4}), with $H^\pr$ of the form of Hamiltonian of $N$
uncoupled stationary oscillators (in terms of old variables),
\beq\lb{o7}
H^\pr  = {1\over 2}\sum_{k}^{N}\lt[{1\over
m_{k}}p_k^{2}+ m_{k}\omega^{2}_{k}q_k^{2}\rt] \equiv
H_{\rm ho}(p,q),
\eeq
The target Hamiltonian $H^\pr $ may also be taken as a sum of  stationary
oscillators $H_{\rm ho}$ in terms of the intermediate variables
$q^{(1)}_{k},\,p^{(1)}_{k}$ as well. In the latter case the second CT
$(q^{(1)}_{k},\, p^{(1)}_{k}) \rar  (q_k^\pr, p_k^\pr)$, generated by
$U_{2}(t) = \exp[-(i/\hbar)H_{\rm ho}(q^{(1)},p^{(1)})t] = S_{\rm ho}(t)$,
takes the explicit form of rotations

\beq\lb{o10ab}
\bt{l}
$q_k^\pr  = q^{(1)}_{k}\cos (\omega _{k}t) + \frac{1}{m_{k}\omega
_{k}}p^{(1)}_{k}\sin (\omega _{k}t)$,\\[2mm]
$p_k^\pr  = -m_{k}\omega _{k}q^{(1)}_{k}\sin(\omega_{k}t) +
p^{(1)}_{k}\cos(\omega_{k}t)$.
\et \eeq

Let us briefly elucidate the two CT involved into the proposition 1.
The first one, generated by  $U_1= U_{0}S^\dg(t)$,  brings $H$ to zero,
therefore the  new  states $|\Psi \ra_{1}$  are  time-independent.
This is because $S^\dg(t)$ is an evolution  operator for the
$\c S$ backward in time. After the first CT (generated by $U_1$) the new
canonical variables
$q^{(1)}_{k}= U_{1}(t)q_{k}U^{\dg}_{1}(t)\quad {\rm and}\quad
p^{(1)}_{j}= U_{1}(t)p_{j}U^{\dg}_{1}(t)$
obey  the equations ($\pd U_1/\pd t = iU_{1}H$, $\pd U_1^\dg/\pd t =
-iHU_1^{\dg}$)

\beq\lb{o8b}
\frac{\pd q^{(1)}_{k}}{\pd t} = {i\over \hbar}[U_1HU^\dg_1,q^{(1)}_{k}],
\quad
\frac{\pd p^{(1)}_{k}}{\pd t} = {i\over \hbar} [U_1HU^\dg_1,p^{(1)}_{k}],
\eeq
i.e., $q^{(1)}_k,\,p^{(1)}_k$ are Heisenberg operators for the old system
$\c S $.

The generator of the second CT $U_{2}(t)=S^\pr(t)$ is  recognized as the
evolution operator forward in time for the target system $\c S^\pr$.  In
the construction (\ref{o4}) $U_2$ is applied to the intermediate
Hamiltonian $H_1$.

It is worth noting at the point the case of CT in the system $\c S$,
generated by its own evolution operator $S(t)$. This CT converts $H(t)$
into Hamiltonian $H^{\pr\pr}(t) = S(t)H(t)S^{\dg}(t) + H(t)$.  If $H$ is
time-independent then $S(t)HS^{\dg}(t) = H$ and $H^{\pr\pr} = 2H$. From
$\la\Psi(t)|A|\Psi(t)\ra = \la\Psi|S^\dg(t)AS(t)|\Psi\ra$ we derive that
the new canonical variables in this case,
\beq\lb{o9a}
q^{\pr\pr}_k = S(t)q_{k}S^{\dg}(t)\equiv q^0_k,\quad
p^{\pr\pr}_k = S(t)p_{k}S^{\dg}(t) \equiv p^0_k,
\eeq
when expressed in terms of the old ones, $q_{k},\,p_{j}$, are {\it
integrals of motion} of $\c S$, satisfying the  eq. (\ref{3a}).  Such
integrals  of  motions  for  quadratic systems $H(t)$ have been
constructed in \ci{MMT1} and intensively  used  later \ci{MM,DMM,Selez,T99}.

Consider the symmetry of $H$ under CT. We want to specify the set of CT
for which $H^\pr$, defined in (\ref{o4}), coincides with $H$, i.e. we look
for CT that keep  $H$ invariant (and thus keep the Schr\"odinger equation
invariant),
\beq\lb{o11}
H^\pr \equiv U(t)HU^\dg(t) -i\hbar U(t){\pd U^\dg(t) \over\pd t}  = H.
\eeq
{} For time-independent $U$ eq. (\ref{o11}) reduces to $H = UHU^\dg$.
From (\ref{o4}) For $H^\pr=H$ the CT generator is (see (\ref{o4})) $U(t)
 = S(t)U_0S^\dg(t)$, where $U_0$ is arbitrary unitary operator. Then $\pd
U^\dg(t)/\pd t = (i/\hbar)[U^\dg(t),H]$ and we see that the equality in
(\ref{o11}) is identically satisfied.  Thus, the CT generators $U(t)$ for
which $H^\pr=H$ have the form $S(t)U_0S^\dg(t)$, i.e. $U(t)$ are integrals
of motion for the system: $ [U(t),D(t)]=0$, where $D(t)$ is the
Schr\"odinger operator, $D(t)=i\hbar\pd/\pd t - H$.  In the first  paper
of  refs. \ci{MMT2} the dynamical symmetry group of a system $\c S$ has
been defined as a group of unitary operators, that commute with $D(t)$ and
act irreducibly in the Hilbert space. Now we see that this symmetry group
leaves $H^\pr=H$ and is highly nonunique, since the unitary operator $U_0$
in $U(t)$ is arbitrary -- one can take $U_0$ from irreducible
representations of any Lie group.  Then the set of invariants
$S(t)U_0S^\dg(t)$ realize an equivalent representation of the same group.
For example, by means of the invariants $q^0_{k}$  and $p^0_{k}$ one can
construct  an irreducible representation of the Lie algebra of  the
Heisenberg-Weyl group $H_W(N)$ and the quasi unitary group $SU(N,1)$ as
well \ci{MMT2}.  This  means that the groups $H_W(N)$ and $SU(N,1)$ can
be considered on equal as dynamical symmetry groups of any
$N$--dimensional system.

In the next section we consider the above  described  unitary (canonical)
equivalence approach  in greater detail for quadratic quantum systems, for
which some  explicit solutions can be obtained.

\section{Canonical transformations of quadratic systems and diagonalization.}

We consider the general $N$--dimensional  nonstationary  quantum system
with Hamiltonian $H(t)$, that is a homogeneous quadratic form of
coordinates and moments,

\beq\lb{o12a}
H(t) = \c A_{jk}(t)p_{j}p_{k}+ \c B_{jk}(t)p_{j}q_{k}+ \td\c B
_{jk}(t)q_{j}p_{k} + \c C_{jk}(t)q_{j}q_{k},
\eeq
 where the coefficients $\c A_{jk}(t) = \c A_{kj}(t)$, $\c
B_{jk}(t),\,\td\c B_{jk}(t)$   and $\c C_{jk}(t) =  \c C_{kj}(t)$ are
arbitrary functions of time. From $H^\dg = H$ it follows that $\c
A_{jk}(t)$ and $\c C_{jk}(t)$ are real, and $\c B_{jk}(t) =
\td\c B_{kj}^*(t)$. It is not a significant restriction to take $\c
B_{jk}$ real and put $\c B_{jk}(t) = \td\c B_{kj}(t)$ (the imaginary parts
of $\c B_{jk}$ can be eliminated by adding a non-operator term to $H$).
In (\ref{o12a}) the summation over the repeated  indices is adopted. We can
introduce $N$--component vectors $\vec{q} = (q_{1},q_{2},\ld ,q_{N}),\quad
\vec{p}= (p_{1},p_{2},\ld ,p_{N}), $ and $N\times N$ real matrices $\c
A(t)$, $\c B(t)$, $\c C(t)$\, ($\c A(t)$ and  $\c C(t)$  are  symmetric)
and rewrite the Hamiltonian (\ref{o12a}) in a more compact form

$$H(t) = \vec{p}\c A(t)\vec{p} + \vec{p}\c B(t)\vec{q} +
\vec{q}\c B(t)^T\vec{p} + \vec{q}\c C(t)\vec{q}, $$
where $\c B^{T}$ is the transposed of $\c B$. To shorthand the notations
it is convenient to introduce the $2N$--vector $\vec{Q} =
(\vec{p},\vec{q})$ and $2N\times 2N$ matrix $\c H$ (the grand matrix) and
rewrite the Hamiltonian (\ref{o12a}) as ($\mu ,\nu  = 1,2,
\ld, 2N$)
\beq\lb{o12b}
H(t) = \c H_{\mu \nu }Q_{\mu }Q_{\nu } \equiv \vec{Q}\c
H(t)\vec{Q},\qquad
\c H = \matb[{\c A} {\c B} {\c B^{T}} {\c C}]\, .
\eeq

We note that nonhomogeneous quadratic Hamiltonians  (i.e., Hamiltonians of
the form (\ref{o12a}), (\ref{o12b}) with linear terms added)  can be
easily reduced to the forms (\ref{o12a}), (\ref{o12b}) by means of
simple time-dependent displacement transformations.

Let $H^\pr $ be an other quadratic Hamiltonian
\beq\lb{o13}
H^\pr (t) =\vec{Q}\c H^\pr (t)\vec{Q},\qquad \c H^\pr  =
\matb[{\c A^\pr} {\c B^\pr} {\c B^\pr{}^{T}} {\c C^\pr}].
\eeq
Then the unitary operator $U(t)$, eq. (\ref{o4}), which relates
canonically  Hamiltonians (\ref{o12b}) and (\ref{o13}), is an exponent of
a quadratic in $\vec{q}$ and $\vec{p}$ form (we take $U_0\in Mp(N,R)$),

\beq\lb{o14}
U(t) = S^\pr(t)U_0S^\dg(t) = \exp\lt[{i\over \hbar }\vec{Q}\td{\c H}(t)
\vec{Q}\rt],
\eeq
 where $\tilde{\c H}(t)$ is a new grand matrix of the form (\ref{o12b})
and (\ref{o13}). $\tilde{\c H}(t)$  can be expressed in terms of the
Hamiltonian matrices $\c H(t)$ and $\c H^\pr (t)$ using the
Baker-Campbell-Hausdorff formula. In this case the  operator  (\ref{o14})
generates linear transformation of coordinates and moments  (we  write it
in $N\times N$ and $2N\times 2N$ matrix forms),
\beq\lb{o15}
\vec{Q}^\pr = \Lam(t)\vec{Q}\quad {\rm or}\quad \mata[{\vec{p}\,^\pr}
{\vec{q}\,^\pr}] =
\matb[{\lam_{pp}} {\lam_{pq}} {\lam_{qp}} {\lam_{qq}}]\mata[{\vec{p}}
{\vec{q}}],
\eeq
where $\lam_{pp}, \lam_{pq}, \lam_{qp} \hbox{ and } \lam_{qq}$ are
$N\times N$ submatrices of $\Lam(t)$.

{}From eqs. (\ref{H'}), (\ref{o12b}), (\ref{o13}) and (\ref{o15}) we
obtain the following relation between the symmetric matrices $\c H$, $\c
H^\pr$ and $\tilde{\c H}$ (\ref{o14}) and the symplectic matrix $\Lam $,

\beq\lb{o16}
{d\over dt}\tilde{\c H}(t) = -\c H^\pr (t) + \Lam ^{T}\c H(t)\Lam .
\eeq
 We see that for a given $\tilde{\c H}(t)$ and $\c H(t)$
this is a simple linear equation for $\c H^\pr (t)$. However for a
given Hamiltonian matrices $\c H$ and $\c H^\pr $  this  is
highly nonlinear equation for $\tilde{\c H}(t)$ since the  matrix
$\Lam (t)$  is  to  be expressed in terms of $\tilde{\c H}(t)$
again: $\Lam \vec{Q}= U(t)\vec{Q}U^{\dg}(t)$.  Nevertheless  for any given
(differentiable with respect to $t$) matrices $\c H(t)$  and $\c
H^\pr (t))$ and for a given  initial  condition $\tilde{\c H}_{0}$
the  above system of equations  has  unique solution for $\tilde{\c H}(t)$,
since the expression of $\Lam $ in terms  of $\tilde{\c H}$  is
also differentiable and Peano theorem could be applied \ci{Kamke}.

In this scheme $\Lam (t)$ is naturally represented as a  product  of
two other $2N\times 2N$ matrices $\Lam ^{(1)}$ and $\Lam ^{(2)}$ of
the form (\ref{o15})  corresponding  to the two successive CT generated by
$U_1(t)$ and $U_2(t)$:
\beq\lb{o17}
\Lam  = \Lam ^{(2)}\Lam ^{(1)};\quad \vec{Q}^{(1)} =
\Lam^{(1)}\vec{Q},\quad \vec{Q}^\pr  = \Lam ^{(2)}\vec{Q}^{(1)}.
\eeq
The matrices $\Lam ^{(1)}$ and $\Lam ^{(2)}$ are seen to
be solutions of  the  first  order  linear equations,
\beq\lb{o18}
{d\over dt}\Lam ^{(1)}= \Lam^{(1)}F^{(1)}(t),\quad {d\over
dt}\Lam ^{(2)}= F^{(2)}(t)\Lam^{(2)},
\eeq
where
\beq\lb{o18a}
F^{(1)}(t) = -2J\c H(t), \quad F^{(2)}(t) = 2J\c H^\pr (t),\qquad J =
\matb[0 1 {-1} 0]\, .
\eeq
 If $H^\pr$ is diagonal as for the oscillator system (\ref{o7})  then the
second  eq. (\ref{o18})  is  easily solved: $\Lam^{(2)}(t) = \exp(2J\c
H_{\rm ho} t)\Lam^{(2)}_0$.  To perform the diagonalization of a quadratic
$H$ one has also to solve the first matrix equation in (\ref{o18}) and
obtain $\Lam ^{(1)}(t)$, which in principle is always possible.  In  the
case  of stationary initial $H$ the $\tilde{T}$ exponent becomes ordinary
one,  so the  explicit solution  is given by the matrix exponent
$\Lam^{(1)}_0\exp(-2J\c H t)$. So for stationary $H$ the total $\Lambda$
matrix takes the form
\beq\lb{o18c}
\Lam(t) = \exp(2J\c H_{\rm ho} t)\Lam^{(2)}_0\Lam^{(1)}_0\exp(-2J\c
Ht),
\eeq
where $\Lam^{(i)}_0$ are arbitrary symplectic matrices.  One can put
$\Lam^{(1,2)}_0=1$, which corresponds to $U_0=1$ in eq. (\ref{o4}).  
Having obtained explicitly $\Lam (t)$ one can next try to solve eq.
(\ref{o16}) and obtain the generating operator $U(t)$ in the form of the
quadratic exponent (\ref{o14}).

Note, the resulting $H^\pr $ is diagonal in the variables, which we choose
for $H_{\rm ho}$. Let those variables be $p^{(1)}_k,\,q^{(1)}_k$. Then the
final variables $p^\pr_k,\, q^\pr_k$ obey eqs. (\ref{o10ab}). Inverting
the transformations (\ref{o10ab}) we obtain $H^\pr$ diagonal in terms of
the final variables as well: $H^\pr = H_{\rm
ho}(\vec{p}\,^\pr,\vec{q}\,^\pr)$.  In this way we perform explicitly the
first kind diagonalization. If $H^\pr = H_{\rm ho}$ in terms of old
variables $p_k,\, q_k$ (second kind diagonalization), then  $H^\pr$ is
evidently not diagonal in terms of $p^\pr_k,\, q^\pr_k$.

{}For some time-dependent $H(t)$ explicit solutions of eqs. (\ref{o18})
can also be found.  Thus, in the case of $N=1$, following the scheme of
refs. \ci{MMT1,MMT2}, one can express matrix elements of $\Lam^{(1)}(t)$
in terms of a complex function $z(t)$, that obeys the equation of
classical oscillator $\ddot{z} + \Omega^2(t)z =0$, where $\Omega^2(t)$ is
simply determined by the parameters $\c A,\,\c B,\,\, \c C$  of the
Hamiltonian (\ref{o12a}) (for $N=1$ these are not matrices, therefore we
put $\c A=a,\,\c B=b,\,\, \c C=c$),

$$\Omega^2(t) = 4ac+ 2b\dot{a}/a + \ddot{a}/2a - 3\dot{a}^2/4a^2 -
4b^2 - 2\dot{b}.$$
For harmonic oscillator with varying frequency $\omega(t)$ we have
$\Omega^2(t) = \omega^2(t)$. It is seen that an $\Omega(t)$ corresponds to
a class of quadratic $H(t)$. For example constant $\Omega$ corresponds to
the stationary oscillator and to the oscillators with varying mass (damped
oscillators) $m(t) = m_0\exp(-2bt)$ and $m(t)=m_0\cos^2bt$, considered
later by many authors (see refs. in \ci{MM,DMM,T99}).  Analytical
solutions to the equation of $z(t)$ are known for a variety of
"frequencies" $\Omega(t)$. In the case of an oscillator with varying
frequency the diagonalizing CT generator $U(t)$ has been expressed in
terms of $z(t)$ in \ci{Selez}.

Let us briefly  discuss  the  algebraic  properties  of  the  matrix
$\Lam(t)$ and its submatrices $\lam_{pp}, \lam_{pq}, \lam_{qp}$, and
$\lam_{qq}$.
From the canonical commutation relations it follows that  $\Lam(t)$ obeys
the relation (the symplectic conditions, $J$ defined in eq. (\ref{o18a}))
\beq\lb{o19a}
\Lam J \Lam ^T= J,
\eeq
 which for the $N\times N$ matrices $\lambda _{qq}, \lambda
_{pp}, \lambda _{qp}$ and $\lambda _{pq}$, defined in eq. (\ref{o15})),  read

\beq\lb{ccr-cond}  
\lam_{pp}\lam_{qq}^T - \lam_{pq}\lam_{qp}^T = 1,\qquad
\lam_{qq}\lam_{qp}^T = \lam_{qp}\lam_{qq}^T,\quad \lam_{pq}\lam_{pp}^T
= \lam_{pp}\lam_{pq}^T.
\eeq
The set of matrices that obey the relation (\ref{o19a}) is  defined as the
symplectic matrix group $Sp(N,R)$ (the transformation
$\vec{x}\,^\pr=\Lam^T\vec{x}$ preserves the quadratic form
$\vec{x}J\vec{x}$).  It has $N(2N+1)$ real  parameters.  The rank of its
Lie algebra is $N$ (following \ci{BR} we use the notation $Sp(N,R)$
instead of $Sp(2N,R)$).\, It is known that in classical mechanics the set
of linear homogeneous CT forms  a symplectic group $Sp(N,R)$.  
In  the quantum  case  the  set  of  matrices $\Lam $,  that  realize
homogeneous linear transformations of the operators of coordinates and
moments close the same group.  However  the  set  of unitary operators $U$
for which $U\vec{q}\,U^{\dg}$ and $U\vec{p}\,U^{\dg}$ are linear
combinations of $\vec{p}$ and $\vec{q}$ contains one extra parameter,
namely the phase factor.  If  one considers CT in  greater  detail  as
transformations of coordinates, moments and {\it vectors} in Hilbert
space  one  has  to count  the  phase factors as well and then we  get
the  larger  group $Sp(N,R)\times U(1) \equiv \tilde{Mp}(N,R)$. If we
consider transformations of coordinates,  moments and {\it states} we have
to factorize over $U(1)$: $\tilde{Mp}(N,R)/U(1) = Mp(N,R)$.  The resulting
group $Mp(N,R)$ is called {\it methaplectic} group.  It  is double
covering of $Sp(N,R)$.  The  Lie algebras of $Mp(N,R)$ and $Sp(N,R)$ are
isomorphic \ci{BR,Yeh}.  They  are  of dimensions $N(2N+1)$ and  this  is
the number  of independent  matrix elements of matrix $\tilde{\c H}$ in
(\ref{o14}). The generators $U(t)$ of linear CT (\ref{o15}) can be
considered as operators of  the unitary  (but  not  faithful)
representation $U(\Lam )$ of the symplectic group $Sp(N,R)$. One can use
the group representation technique
\ci{BR} to represent $U(t)\in Sp(N,R)$ in several factorized forms.  In
the case of one dimensional nonstationary harmonic oscillator the
diagonalizing CT generator $U(t)$, $U(t)\in SU(1,1)$, and its factorized
forms have been considered in \ci{Selez}.

If one considers Hamiltonians  (\ref{o12a})  with  linear  terms
$\vec{d}(t)\vec{p} +  \vec{e}(t)\vec{q}$  added, then  in  the  same  way
one  would  get  that  such inhomogeneous quadratic Hamiltonians can be
diagonalized to  the  form (\ref{o7}) by means of the same $U(t)$, eq.
(\ref{o4}), this time $U(t)$ being an element of the semidirect product
group $Mp(N,R)\times\!\!\!\!\!\supset H_{w}(N)$, where $H_{w}(N)$  is  the
$N$ dimensional  Heisenberg--Weyl group.

\section{Diagonalization of uncertainty matrix and
 minimization of characteristic inequalities}
\medskip

The established possibility  of  converting (by means of time-dependent
CT) any $N$--dimensional Hamiltonian $H$  to that of the system of
uncoupled harmonic oscillators suggests to expect  that the dispersion
matrix $\mbox{\lg$\sig$}(\vec{Q},\rho )$ of canonical observables
$Q_\nu$\,, $\nu =1,\ld 2N,$ in any (generally mixed) quantum state $\rho $
could be diagonalized by means of some  state dependent  CT.  It turns out
that this really holds \ci{SCB,T95}.

Let us recall the notion of dispersion matrix
$\mbox{\lg$\sig$}(\vec{X},\rho)$ (called also fluctuation matrix, or
uncertainty matrix). This is an $n\times n$ matrix constructed by means
of  the  second  moments (the variances and covariances) of observables
$X_1,\ld,X_n$ in a state $\rho$.  The matrix elements
$\mbox{\lg$\sig$}_{\mu\nu}$ of $\mbox{\lg$\sig$}$ are defined as
covariances $\Dlt X_\mu X_\nu$ of the observables $X_\mu$ and $X_\nu$,
$\nu, \mu = 1,\ld,n$, 
$$
\mbox{\lg$\sig$}_{\mu \nu }(\vec{X},\rho) = {1\over 2}\la X_{\mu}
X_{\nu}+ X_{\nu}X_{\mu}\ra - \la X_{\mu}\ra \la X_{\nu}\ra \equiv
\Dlt X_\mu X_\nu(\rho).
$$
The  matrix $\mbox{\lg$\sig$}(\vec{X},\rho)$ is symmetric by construction.
It satisfy the {\it characteristic uncertainty relations} \ci{TD}\,\,
$C_r^{(n)}(\sig(\vec{X},\rho)) \,\geq\, C_r^{(n)}(C(\vec{X},\rho))$,\,\,
where $C(\vec{X},\rho)$ is the $n\times n$ antisymmetric matrix of the
means of commutators of $X_\mu$ and $X_\nu$,\, $C_{\mu\nu} =
-i\la[X_\mu,X_\nu]\ra/2$,\, and $C_r^{(n)}(M)$, $r=1,\ld,n$, are the
characteristic coefficients of a $n\times n$ matrix $M$ \ci{Gant}. The
characteristic coefficient of maximal order $r=n$ is the determinant of
$M$.  The characteristic uncertainty relation of maximal order $r=n$,
\beq\lb{RUR}
\det\mbox{\lg$\sig$}(\vec{X},\rho) \geq \det\,C(\vec{X},\rho),
\eeq
has been established by Robertson \ci{R34} and is called {\it Robertson
uncertainty relation}.
For $N=2$ inequality (\ref{RUR}) recovers the Schr\"odinger
uncertainty relation \ci{S30},
$(\Dlt X)^2(\Dlt Y)^2 - (\Dlt XY)^2 \geq \lt|\la[X,Y]\ra\rt|^2/4$,
which for the canonical pair $q,p$,\,\, $[q,p]=i$,\, takes the simpler
form of\, (hereafter we put $\hbar=1$)
\beq\lb{qpSUR}
(\Dlt p)^2(\Dlt q)^2 - (\Dlt pq)^2 \geq 1/4.
\eeq
The proof of (\ref{RUR}) is based on the nonnegativity of the matrix $R =
\mbox{\lg$\sig$} +iC$ \ci{R34}.  Properties of $R$ (to be called Robertson
matrix) are reviewed in \ci{T97}. Here we need the nonnegativity property
of $\mbox{\lg$\sig$}(\vec{X},\rho)$.\\

{\sm \bf Proposition 2.} {\it The uncertainty matrix for any $n$ observables
$X_1,\ld,X_n$ is nonnegative definite},
$\mbox{\lg$\sig$}(\vec{X},\rho) \geq 0$.\\[-3mm]

{\it Proof}. The proof relies to the Robertson inequality (\ref{RUR}) and
on the observation that the principal submatrices
$m(X_{i_1},\ld,X_{i_r},\rho)$, $r \leq n$, of $\mbox{\lg$\sig$}$ can be
regarded as uncertainty matrices for $r$ observables $X_{i_1},\ld,X_{i_r}$
in the same state $\rho$. Therefore the submatrices
$m(X_{i_1},\ld,X_{i_r}),\rho)$ also satisfy Robertson relation
(\ref{RUR}), i.e. their determinants (the principal minors of
$\mbox{\lg$\sig$}$) are nonnegative.  And if all principal minors of a
matrix $M$ are nonnegative, then $M\geq 0$ \ci{Gant}. End of the proof.\\

The uncertainty matrix $\mbox{\lg$\sig$}(\vec{Q},\rho)$ for canonical
observables $Q_\mu$, $\mu = 1,\ld, 2N$: $Q_{k} = p_{k}, Q_{N+k} =
q_{k}$, $k=1,\ld,N$, possess some further properties. \\

{\sm \bf Proposition 3.} {\it The uncertainty matrix for $2N$ canonical
observables $Q_\mu$ is positive definite},
$\mbox{\lg$\sig$}(\vec{Q},\rho) \,>\, 0.$\\[-3mm]

{\it Proof}. From the canonical commutation relations
$[q_k,p_j]=i\dlt_{kj}$ it follows that $\det C(\vec{Q},\rho) = (1/4)^N$.
Then (\ref{RUR}) yields
\beq\lb{qpRUR}
\det\mbox{\lg$\sig$}(\vec{Q},\rho) \geq (1/4)^N.
\eeq
As a symmetric matrix $\mbox{\lg$\sig$}(\vec{Q},\rho)$ can be diagonalized
by an orthogonal transformation $Q_\mu\rar Q_\mu^\pr =
\gam_{\mu\nu}Q_\nu$. The uncertainty matrix for new observables
$Q^\pr_\mu$ is $\mbox{\lg$\sig$}^\pr \equiv
\mbox{\lg$\sig$}(\vec{Q}^\pr,\rho) =  \gam\mbox{\lg$\sig$}\gam^{T}$.  This
transformation preserves the determinant (and all the other characteristic
coefficients) of $\mbox{\lg$\sig$}$. In view of
$\det\mbox{\lg$\sig$}(\vec{Q},\rho) >0$ all diagonal elements of
$\mbox{\lg$\sig$}^\pr$ are positive. 
Therefore $\mbox{\lg$\sig$}(\vec{Q},\rho) > 0$. End of the proof. \\

The desired diagonalization of $\mbox{\lg$\sig$}(\vec{Q},\rho)$ using {\it
linear canonical transformations} \ci{SCB,T95} now follows from the
Proposition 3 and the known theorem \ci{Will,Tiko,Bogdan} that any positive
definite symmetric matrix $M$ can be diagonalized by means of congruent
transformation with a symplectic matrix $\Lam$,  $M\rar M^\pr=\Lam
M\Lam^T$.  In \ci{SCB} the diagonalization of
$\mbox{\lg$\sig$}(\vec{Q},\rho)$ is performed explicitly by means of three
consecutive linear canonical transformations. The diagonal elements
$s_\mu$ of the diagonalized $\mbox{\lg$\sig$}^\pr$ are variances $(\Dlt
Q_\mu^\pr)^2$.  Additional scaling transformations $q_i^{\pr} \rar
q_i^{\pr\pr}=\alf_iq_i^\pr$ with $\alf_i = (\Dlt p_i^\pr/\Dlt q_i^\pr)^{1/2}$
equalize the variances of $q_i^{\pr\pr}$ and $p_i^{\pr\pr}
=p_i^\pr/\alf_i$.  Note that: (a) the diagonalizing symplectic matrix
$\Lam$ is not unique \ci{Tiko,Yeh};\,\, (b) $\Lam$ is state-dependent.
Therefore it may depend on time when the state is time-dependent.

Denoting the generator of the total diagonalizing canonical transformation
by $U(\Lam)$ [$Q^\pr = U(\Lam)QU^\dg(\Lam)$, $U(\Lam)\in Mp(N,R)$] we
obtain the equality $\mbox{\lg$\sig$}(\vec{Q},\rho^\pr) =
\mbox{\lg$\sig$}(\vec{Q}^\pr,\rho)$, where $\rho^\pr=U\rho U^\dg$. Thus every
state $\rho$ is {\it unitary and methaplectically equivalent} to a
state $\rho^\pr$, in which the uncertainty matrix
$\mbox{\lg$\sig$}(\vec{Q},\rho^\pr)$ is diagonal with equal variances of
coordinates $q_i$ and moments $p_i$: $\Dlt q_i = \Dlt p_i$.  If the
initial state $|\Psi\ra$ is pure time-dependent state of system $\c S$
with Hamiltonian $H$, then the CT is time-dependent and the new state
$|\Psi^\pr\ra$  obey the Schr\"odinger equation with new Hamiltonian
(\ref{H'}).

Examples of pure states with diagonal uncertainty matrix with equal
variances of coordinates and moments are Glauber multimode coherent states
$|\vec{\alf}\ra$ and  multimode Fock states $|\vec{n}\ra$. Therefore in
the Klauder--Perelomov $Mp(2,R)$ CS $|g,\vec{\alf}\ra =
U(g)|\vec{\alf}\ra$ and $|g,\vec{n}\ra = U(g)|\vec{n}\ra$ ($g$ being the
group element) the dispersion matrices
$\mbox{\lg$\sig$}(\vec{Q}^\pr,g,\vec{\alf})$ and
$\mbox{\lg$\sig$}(\vec{Q}^\pr,g,\vec{n})$ are diagonal and with equal
variances of $q_i^\pr = U^\dg(g)q_iU(g)$ and $p_i^\pr=U^\dg(g)p_iU(g)$. In
$|\vec{\alf}\ra$ {\it all variances are equal and minimal}, $\Dlt q_i =
\Dlt p_i = 1/\sqrt{2}$, whereas in Fock states the variances are equal in
pairs, $(\Dlt q_i)^2 = (\Dlt p_i)^2 = 1/2 + n_i$.  Multimode CS
$|\vec{\alf}\ra$ minimize Robertson inequality (\ref{qpRUR}), whereas in
$|\vec{n}\ra$ one has $\det\mbox{\lg$\sig$}(\vec{Q},\vec{n}) = 
\prod_i(1/2+n_i)$.

It is clear from the above consideration that the uncertainty matrix in
any group-related CS $T(g)|\Psi_0\ra$ with reference vector $|\Psi_0\ra$
equal to $|\vec{\alf}\ra$ or $|\vec{n}\ra$ is diagonalized by the CT
$Q^\pr_\mu = T^\dg(g)Q_\mu T(g)$, that is linear for the group $Mp(N,R)$
only. In physical literature the $Mp(N,R)$ group-related CS
$U(g)|\vec{\alf}\ra$ and $U(g)|\vec{n}\ra$ are known as multimode {\it
squeezed} CS and squeezed Fock states respectively.   The operator
$U(g)\in Mp(N,R)$ can be called multimode squeeze operator \ci{Ma}, its
canonical form being $\exp[(\vec{a}\,^\dg z\vec{a}\,^\dg -
\vec{a}z^*\vec{a})/2]$, where $\vec{a}\,^\dg z\vec{a}\,^\dg = a_i^\dg
z_{ij}a_j^\dg,\quad i,j=1,\ld,N$ \ci{Ma}.  It is more adequate to call it
{\it squeeze and correlation operator} since, e.g., for pure imaginary
$z_{ii}$ it generates covariances of $p_i$ and $q_i$ and doesn't squeeze,
while for real $z_{ii}$ it generates squeezing and doesn't correlate.  The
wave function $\la\vec{x}|U(g)|\vec{\alf}\ra$ of $Mp(N,R)$ CS is Gaussian
(an exponent of $N$--dimensional quadratic form), thereby that states are
also called Gaussian pure states \ci{Littl,Simon}.

It is interesting to note that the multimode squeezed states
$U(g)|\vec{\alf}\ra$ are the unique states to minimize the Robertson
inequality (\ref{qpRUR}).\\

{\sm\bf Proposition 4}. {\it The equality in the multimode Robertson
uncertainty relation, eq.(27), holds in the multimode squeezed states
$U(g)|\vec{\alf}\ra$ ($g\in Mp(N,R)$)  only.}\\

{\it Proof}.  Let $\Lam(\rho)$ be a symplectic matrix that diagonalizes
the  dispersion  matrix $\mbox{\lg$\sig$}(\vec{Q},\rho )$, and $U = U(\Lam)$
-- the generator of the diagonalizing CT $\vec{Q}^\pr = \Lam(\rho)\vec{Q}
= U(\Lam)\vec{Q}U^\dg(\Lam)$.  $U(\Lam)$ belongs to $Mp(N,R)$.  We have

\beq\lb{o33a}
\mbox{\lg$\sig$}(\vec{Q}^\pr,\rho ) = \Lam(\rho)
\mbox{\lg$\sig$}(\vec{Q},\rho )
\Lam ^{T}(\rho) =
\mbox{\lg$\sig$}(\vec{Q},\rho ^\pr ),\quad \rho^\pr  = U(\Lam)\rho U^\dg(\Lam),
\eeq

\beq\lb{o33b}
\mbox{\lg$\sig$}(\vec{Q},\rho^\pr ) = {\rm diag}\{s_{1},\, s_{2},\,\ld,\,
s_{2N}\},
\eeq
where the diagonal elements $s_{\nu }$ are the variances of $q_k$ and
$p_k$ in the new state $\rho ^\pr$:
$s_k = (\Dlt p_k(\rho^\pr))^2$, $s_{N+k}(\rho^\pr) = (\Dlt q_k(\rho^\pr))^2$.
The determinant of $\mbox{\lg$\sig$}(\vec{Q},\rho^\pr)$ is a  product  of
all diagonal elements $s_{\nu}$, $\nu = 1,\ld,2N$,

\beq\lb{o35}
\det\mbox{\lg$\sig$}(\vec{Q},\rho) =\det\mbox{\lg$\sig$}(\vec{Q},\rho^\pr) =
[s_{1}s_{N+1}][s_{2}s_{N+2}] \ld [s_{N}s_{2N}].
\eeq
From Heisenberg uncertainty relation we have for every factor in eq.
(\ref{o35}) the inequality
\beq\lb{qpHURs}  %
s_{k}s_{N+k} = (\Dlt p_k)^2(\Dlt q_k)^2 \ge  1/4.
\eeq
From eqs. (\ref{qpHURs}) and (\ref{o35}) we derive that the
equality  in Robertson relation (\ref{qpRUR}) holds iff the equality in eq.
(\ref{qpHURs}) holds for all modes (for every $k = 1,\ld N)$. The minimal
value of $1/4$ of the product of variances of $q$ and $p$ cannot be
reached in mixed state \ci{SCB} -- it is reached in the Stoler states
\ci{SS} $|\alf,r\ra = \exp[r(a^{\dg 2} - a^2)]\,|\alf\ra$ only (see proof
in the Appendix). Thus the equality in (\ref{qpRUR}) holds in pure states
$U(\Lam)\prod_k S(r_k)|\vec{\alf}\ra$ only.  The unitary operator
$S(r)=\exp[r(a^{\dg 2} - a^2)/2]$ (the squeeze operator) belongs to
$U(1,1) \sim Mp(1,R)$. Therefore the unitary operator $U(\Lam)\prod_k
S(r_k)=U(g)$ belongs to $Mp(N,R)$, and the unique minimizing states are
$Mp(N,R)$-group related CS with reference vector $|\vec{\alf}\ra$. End of
the proof.\\

Since Glauber CS $|\vec{\alf}\ra$ are eigenstates of every annihilation
operator $a_k$ (with eigenvalues $\alf_k$, $k=1,\ld,N$), the minimizing
states $U(g)|\vec{\alf}\ra$
are eigenstates of the canonically transformed annihilation operators
$a^\pr_k = U(g)a_kU^\dg(g)$, which are linear combinations of $a_1,\ld,a_N$:
 $a^\pr_k = u_{kj}a_j + v_{kj}a^\dg_j$. Therefore the minimizing states
$U(g)|\vec{\alf}\ra$ (the multimode squeezed states) can be denoted
equivalently as $|\vec{\alf},u,v\ra$. For $v=0$ (and $u=1$) they coincide
with $|\vec{\alf}\ra$.

For quadratic Hamiltonians the time evolution operator $U_{\rm quad}(t)\in
Mp(N,R)$.  Therefore the time evolution of $|\vec{\alf},u,v\ra$ for
quadratic Hamiltonians is stable, i.e., $U(t)|\vec{\alf},u,v\ra =
|\vec{\alf},u(t),v(t)\ra$. The evolved states $|\vec{\alf},u(t),v(t)\ra$
are eigenstates of the new annihilation operators $A_k(t) =
U(t)a^\pr_kU^\dg(t)$, which are again linear in $a_j$ and $a^\dg_j$ and
are integrals of motion of quadratic system.  Overcomplete system of
eigenstates $|\vec{\alf},t\ra$ of integrals of motion $A_k(t)$ has been
constructed in ref. [5] and used  later  in many papers [10].

A further property of the uncertainty matrix  (the fourth one) we want to
note here is referred to its {\it symplectic character}: the normalized
uncertainty matrix $\mbox{\lg$\td{\sig}$} =
\mbox{\lg$\sig$}/\lt(\det\mbox{\lg$\sig$}\rt)^{1/2N}$
is symplectic
for a certain class of states. In order to find out that states we
note the invariance of the symplectic property of a matrix $M$ under the
congruent transformation $\Lam M\Lam^T$ with a symplectic $\Lam$: if $M$
is symplectic, that is $MJM^T=J$, then $M^\pr = \Lam M\Lam^T$ is also
symplectic. This symplectic invariance can be easily proved using the known
property that if
$\Lam J\Lam^T=J$ then one also has $\Lam^TJ\Lam=J$:
$M^\pr JM^{\pr T} = \Lam M\Lam^T J \Lam M^T\Lam^T =
\Lam MJM^T\Lam^T = \Lam J\Lam^T = J$.
This invariance enables us to study the symplectic properties of
$\mbox{\lg$\sig$}$ in its simpler diagonal form.  For diagonal uncertainty
matrix $\mbox{\lg$\td{\sig}$} = {\rm diag}\{s_1,\ld,s_N\}$ the symplectic
condition $\mbox{\lg$\td{\sig}$}\,J\,\mbox{\lg$\td{\sig}^T$} = J$ reduces to
\beq\lb{sycond2}
 s_1s_{N+1} = s_2s_{N+2} = \ld = s_Ns_{2N}
 = \lt(\det\mbox{\lg$\sig$}\rt)^{1/N}.
\eeq
One solution to (\ref{sycond2}) can be immediately pointed out, recalling
the meaning of $s_\mu$ as the variance of $Q_\mu$: the uncertainty
matrix in the multimode Glauber CS $|\vec{\alf}\ra$ is diagonal with
$s_k =
(\Dlt p_k)^2 = 1/2$, $s_{N+k} = (\Dlt p_k)^2 = 1/2$, $k=1,\ld,N$, which
clearly satisfy (\ref{sycond2}).
Therefore the normalized uncertainty matrix in pure states
$U(g)|\vec{\alf}\ra$ that are unitary equivalent to Glauber CS with
$U(g)\in Mp(N,R)$ {\it is symplectic}. These states, as we have already
noted, are called Gaussian pure states or multimode squeezed CS. In fact
the symplectic character of the normalized uncertainty matrix for Gaussian
pure states was established in \ci{Simon}: in that states our
$\mbox{\lg$\td{\sig}$}$ is equal to $2\mbox{\lg$\sig$}$ and this
quantity coincides with the matrix
$\underline{G}(\underline{U}^{-1},-\underline{V})$ of \ci{Simon}, which
was shown to be symplectic \ci{Simon}.

A second solution to (\ref{sycond2}) is provided by the uncertainty matrix
in (multimode) Fock states $|\vec{n}\ra$ with equal numbers $n_k=n$
(equal numbers of photons in every mode). In $|\vec{n}\ra$ we have $(\Dlt
p_k)^2 = 1/2+n_k = (\Dlt q_k)^2$. Therefore in states $U(g)|\vec{n}\ra$
with $n_1=\ld=n_N$ and $U(g)\in Mp(N,R)$ the normalized uncertainty matrix
is symplectic.  The above two families of states do not exhaust the set
states with symplectic (normalized) uncertainty matrix.

Let us write down the symplectic conditions and the Robertson relation for
$\mbox{\lg$\sig$}(\vec{Q},\rho)$ in terms  of  the  four
$N\times N$ blocks $\sig_{pp}(\rho)$, $\sig_{qq}(\rho)$, $\sig_{pq}(\rho)$
and $\sig_{qp}(\rho)$,

\beq\lb{o41}
\mbox{\lg$\sig$}(\vec{Q},\rho) = \matb[{\sig_{pp}(\rho)} {\sig_{pq}(\rho)}
{\sig_{qp}(\rho)} {\sig_{qq}(\rho)}] ,
\eeq
Inserting this into $\mbox{\lg$\td{\sig}$}\,J\,\mbox{\lg$\td{\sig}^T$} = J$,
and taking into account that $\sig_{pp}$ and $\sig_{qq}$ are symmetric,
and $\sig_{pq} = \sig_{qp}^T$ we obtain

\beq\lb{o42a}
\sig_{pp}\sig_{qq} - (\sig_{pq})^2 = \lt(\det\mbox{\lg$\sig$}\rt)^{1/N},
\eeq
\beq\lb{o42b}
\sig_{pp}\sig_{qp} - \sig_{pq}\sig_{pp} = 0,\quad \sig_{qp}\sig_{qq} -
\sig_{qq}\sig_{pq} = 0.
\eeq

Squeezed CS $U(g)|\vec{\alf}\ra$  minimize (\ref{qpRUR}), i.e.
$\det\mbox{\lg$\sig$} = (1/4)^N$. Therefore in $U(g)|\vec{\alf}\ra$ the
symplectic condition (\ref{o42a}) reads $\sig_{pp}\sig_{qq} -
(\sig_{pq})^2 = 1/4$.  The latter formula was obtained in \ci{Ma} for the
squeezed CS of the form
$\exp[(\vec{a}^\dg z\vec{a}^\dg - \vec{a}z^*\vec{a})/2]\,|\vec{\alf}\ra$
by direct calculations (but with no reference to Robertson inequality,
neither to the symplecticity of the uncertainty matrix). In squeezed Fock
states $U(g)|\vec{n}\ra$ we have $\det\mbox{\lg$\sig$}(g,\vec{n}) =
\prod_k(1/2+n_k) \geq (1/4)^N$. For these states the symplectic condition
(\ref{o42a}) is valid iff $n_k=n$, and reads $\sig_{pp}\sig_{qq} -
(\sig_{pq})^2 = (1/2 + n)^2$.

In terms of the $N\times N$ matrices Robertson inequality (\ref{qpRUR})
takes the form (using known formulas for the block matrices \ci{Gant})

\beq\lb{qpRUR2}
\det[\sig_{pp}\sig_{qq} - \sig_{pp}\sig_{qp}\sig_{pp}^{-1}\sig_{pq}] \geq
(1/4)^N.
\eeq
For \mbox{\lg$\td{\sig}$} symplectic we have $\sig_{pp}\sig_{qp} =
\sig_{pq}\sig_{pp}$, and the Robertson relation simplifies to
$\det[\sig_{pp}\sig_{qq} - (\sig_{pq})^2] \geq (1/4)^N$. This form is
quite similar to that of Schr\"odinger inequality (\ref{qpSUR}) for $p$ and
$q$: for $N=1$ we have $\sig_{pq} = \Dlt pq$, $\sig_{pp} = \Dlt pp \equiv
(\Dlt p)^2$, and $\sig_{qq} = \Dlt qq \equiv (\Dlt q)^2$.

It is curious to note that the Robertson matrix $\td{R}$ for normalized
$\mbox{\lg$\td{\sig}$}(\vec{Q},\rho)$ and  $\td{C}(\vec{Q},\rho) =
C(\vec{Q},\rho)/\lt(\det C(\vec{Q},\rho)\rt)^{1/2N}$, $\td{R} =
\mbox{\lg$\td{\sig}$} + i\td{C}$, is also symplectic
for squeezed CS and squeezed Fock states with $n_k=n$: $\td{R}J\td{R}^\dg
= J$, that is $\td{R} \in Sp(N,C)$.\\

\section{Appendix}

{\sm\bf Proposition A1.} {\it  Heisenberg inequality
$(\Dlt q)^2(\Dlt p)^2 \geq 1/4$
is minimized in the Stoler states $|\alf,r\ra = \exp[r(a^{\dg2} -
a^2)/2]\,|\alf\ra$ only}.\\[-3mm]

Let $\rho$ be a general mixed state. Any mixed state can be represented in
the form  $\rho = \sum_k
\rho_k|\psi_k\ra\la\psi_k|$, where $\rho_k \geq 0$, and $\{|\psi_k\ra\}$
is some complete orthonormal set of pure states. The mean value of an
operator $X$ in $\rho$ is given by
$\la X\ra = {\rm Tr}(X\rho)$.
Consider the
mean value of the operator $b^\dg(\lam)b(\lam)$, where
\beq\lb{b}
b(\lam) = \lam q+ip - (\lam\la q\ra + i\la p\ra,\quad \lam \in R.
\eeq
For positive $\lam$ the operator $b(\lam)$ is, up to a factor $1/\sqrt{2\lam}$,
a boson annihilation operator, $[b,b^\dg]=2\lam$, and $b^\dg b$ is, up to
a factor $1/2\lam$, the number operator, which is nonnegative definite.
For negative $\lam$ the operators $b$ and $b^\dg$ are interchanged. The
mean of $b^\dg(\lam)b(\lam)$ in $\rho$ reads
\beq\lb{<bb1>}
\la b^\dg(\lam)b(\lam)\ra =
\sum_k\rho_k\la\psi_k|b^\dg(\lam)b(\lam)|\psi_k\ra,
\eeq
where all means $\la\psi_k|b^\dg(\lam)b(\lam)|\psi_k\ra$ are nonnegative.
On the other hand, by the use of (\ref{b}), this nonnegative mean
$\la b^\dg(\lam)b(\lam)\ra$ can be written as
\beq\lb{<bb2>}
\la b^\dg(\lam)b(\lam)\ra =
\lam^2(\Dlt q)^2 - \lam + (\Dlt p)^2 \geq 0.
\eeq
The $\lam$-roots of the equation $\lam^2(\Dlt q)^2 - \lam + (\Dlt p)^2 =
0$ must be real, wherefrom one deduces Heisenberg inequality. The equality
in Heisenberg relation corresponds to the equality in (\ref{<bb2>}), i.e.
to the vanishing $\la b^\dg(\lam)b(\lam)\ra$. From (\ref{<bb1>}) it is
seen that $\la b^\dg(\lam)b(\lam)\ra = 0$ if and only if
$\la\psi_k|b^\dg(\lam)b(\lam)|\psi_k\ra = 0$ for every $k=1,\ld $ (in view
of $\rho_k \geq 0$).  From the {\it uniqueness of the vacuum state} it
follows that all $\rho_k$ but one (say $\rho_1$) must be zero.  Therefore
$\la b^\dg(\lam)b(\lam)\ra = 0$ in pure state $|\psi\ra$ only and iff it
is an eigenstate of $\lam q+ip$.  The final step is to identify the
minimizing pure state with $|\alf,r\ra$.  The minimizing state must be
eigenstate of $\lam q+ip$ for some real $\lam$.  $|\alf,r\ra$ are
eigenstates of $\lam q +ip)$ with $\lam= [\cosh(2r) -
\sinh(2r)]/[\cosh(2r) + \sinh(2r)]$ and eigenvalue
$\alf\sqrt{2}/[\cosh(2r) + \sinh(2r)]$.  Thus $(\Dlt q)^2(\Dlt p)^2 = 1/4$
holds in states $|\alf,r\ra$ only.  In slightly different notations the
proof of the statement that a state with absolute minimum of the product
$(\Dlt q)^2(\Dlt p)^2$ is a pure state is given in \ci{SCB}.

\end{document}